\documentclass{aastex63}

\newcommand{\fracbrac}[2]{\left(\frac{#1}{#2}\right)}

\newcommand{\zm}[1]{ \textcolor{black}{#1}} 
\newcommand{\change}[1]{ \textcolor{black}{#1}}

\usepackage{natbib}
\usepackage{graphicx}
\usepackage{amsmath}
\usepackage{multirow}
\accepted{APJ}

\shorttitle{Two Body cross sections}
\shortauthors{Murray et al.}

\begin{document}

\title{Numerical Determination of the Gravitational cross sections of an Accreting Binary}

\correspondingauthor{Zachary~Murray}
\email{murray@geoazur.unice.fr}
\author[0000-0002-8076-3854]{Zachary Murray}
\affiliation{Université Côte d'Azur, France }

\begin{abstract}
A significant amount of work has been devoted to the study of small binary solar system objects.  The majority of these binaries, especially among the near-earth or main belt asteroids have small radius ratios, implying a large difference in size between the primary and its companion. Farther from the sun, the binary fraction increases, with the Kuiper Belt having many known binaries with radius ratios of order unity. In this paper, we consider the runaway growth of a binary system in an accretionary stream of small particles.  We \zm{perform brute-force integrations, each with 10 million test particles} and numerically compute \zm{the gravitational} cross sections for each member of the binary as a function of the system's separation and mass ratio.  We show that the behavior of the cross section is complex, and it can be either diminished or enhanced depending on the orbital configuration.  In the regime where gravitational focusing dominates the accretion process, we show that binaries grow towards smaller mass ratios than would be expected given single-body cross sections.  Finally, we provide a grid of these cross sections for use in the future study of such systems. 
\end{abstract}


\section{Introduction} \label{sec:intro}

The solar system is a strongly hierarchical gravitational system.  The sun contains over $99.9 \%$ of the mass of the total system, dwarfing the masses of all the planets. This system of hierarchy continues within individual planetary systems.  Of the major planets that possess satellites five  (Mars, Jupiter, Saturn, Uranus and Neptune) have masses in excess of $4500$ times the mass of their most massive satellite.  The exception is the Earth whose mass still exceeds that of the moon by $\approx 80$ times. Among the small bodies interior to the orbit of Jupiter, known binaries typically follow this same scheme, often having small mass ratios, with similarly massed binaries being relatively rare. The formation of many of these bodies can be explained by some combination of the YORP/BYORP mechanism and tidal processes summarized in \citet{Marchis_2011}. 

However, a few objects in the solar system are outliers and possess much larger mass ratios, with the most notable examples in the inner solar system being the binary asteroid (90) Antiope and binary Jupiter Trojan (617) Patroclus. Both of these bodies are large, with diameters around $100 \mathrm{km}$, and hence are not strongly affected by thermal effects which can drive fission in smaller asteroids \citep{Buie_2015,Walsh_2015}. In addition, collisions rarely produce objects of equal mass \citep{Durda_2004}, hence both of these systems pose challenges to traditional models of binary formation.  

Correspondingly, both objects have been subjected to significant study. (90) Antiope consists of two components, with radii of $43 \mathrm{km}$ and $42 \mathrm{km}$ respectively. These bodies orbit around their common center of mass with a semi-major axis of about $4$ primary radii in a highly inclined $63^\circ$ nearly circular orbit \citep{Descamps_2007}.  The bodies have a low bulk density of $1.25 \; \mathrm{ g cm^{-3}}$ and are tidally synchronized with a correspondingly short tidal dissipation timescale of only $10^4 \mathrm{years}$ \citep{Michalowski_2002, Michalowski_2004}. Component resolved spectroscopy of (90) Antiope shows that the two bodies have nearly identical spectra and therefore likely have similar compositions \citep{Marchis_2011}. Spectra taken by \citet{Hargrove_2015} are suggestive of water ice, consistent with the bodies' low densities.  

The properties of (617) Patroclus are similar to those of (90) Antiope.  Data from \citet{Mueller_2010} suggest the system is fully synchronized and has a bulk mass density of just $1.08 \pm 0.33 \; \mathrm{g cm^3}$, again implying a water ice composition.  The orbit of the two bodies is consistent with being circular \citep{Marchis_2006, Berthier_2007, Berthier_2020} with low inclination. Patroclus, the primary, is roughly $62 \mathrm{km}$ in radius, and Menoetius, the secondary, is about $60 \mathrm{km}$.  (617) Patroclus also forms a tight binary with a semimajor axis of approximately 10 primary radii.   Occultation observations show that the two components are of similar size; their similar magnitudes imply similar albedos and compositions \citep{Buie_2015}.

Due to their low densities, it has been argued that both of these binaries originated in the Kuiper Belt and may have migrated to their present positions in the solar system~\citep{Goldreich_2002,Morbidelli_2005,Nesvorny_2010,Nesvorny_2018}.  Indeed, the outer solar system contains a multitude of similar systems in an abundance that both greatly exceeds those found in the inner systems and possesses more diverse orbital configurations~\citep{Funato_2004}.  Examples among the Centaurs and Kuiper belt objects include (42355) Typhon \citep{Noll_2006}, (65489) Ceto/Phorcys \citep{Grundy_2007} and many others \citep{Noll_2008}.

Currently, a only few dozen Kuiper Belt binaries are known. If the components of these binaries are assumed to have similar albedos and compositions, then the magnitude difference can be transformed into a radius or mass ratio.  The distribution of radius ratios was explored in \citet{Nesvorny_2010}, which argued that direct gravitational collapse can produce binaries with a radius ratio distribution roughly consistent with the observed distribution in the Kuiper belt albeit with a potential excess of small mass ratio binaries.  While other ideas about the formation of the Kuiper Belt binaries have been proposed \citep[e.g.][]{Weidenschilling_2002,Goldreich_2002,Schlichting_2008}, regardless of the formation mechanism, the young binaries experience a period of accretion in which they orbit within a \zm{background} of smaller bodies,  \zm{In this regime, the  radius ratio distribution will depend not only on the geometric cross sections of the bodies, but the \textit{gravitational cross section}, that which takes into account time dependent gravitational field as they orbit each other}.  Future surveys, especially LSST, are poised to discover many more binaries that would further constrain the radius ratio distribution. If the relative cross sections of an accreting binary were also known, inferences could be made about the strength of gravitational focusing and correspondingly the velocity distribution in the natal belt.  Therefore, we concern ourselves with exploring gravitational focusing in the regime of closely spaced binaries. We wish to determine the gravitational cross sections of an accreting primary and secondary body as a function of their mass ratio, semi-major axis, and the incident velocity of the small bodies.  \zm{Absent a fully analytic theory of three body interactions, we compute these cross sections numerically by brute force simulation.}  Such cross sections could be important for future studies of binary growth both in the Kuiper belt and in other contexts.  \zm{In this paper we begin by describing our methods for adimensionalizing the problem and show convergence checks to demonstrate that our numerically computed \change{cross} sections are trustworthy.  We then run our code over a grid of parameters and present our averaged results, we also include a handful of specific cases, to give readers intuition about the dynamics that shape the cross sections in different regimes.  \change{Finally we examine the case of a binary which is increasing in mass via accretion }and compare our numerical cross sections to geometric ones. }
\section{Methods}

    In the traditional, single-body, gravitational focusing model, one considers the motion of a stream of small particles with negligible masses incident on some larger body of mass $M_1$ and radius $R_1$.  The effective cross section of such a body can be related to its escape velocity and the velocity of the incident particles. It can be written as 
    \begin{equation}
        \sigma_{eff} = \sigma_{geo} \left(1 + \frac{V_e^2}{V_i^2} \right) = \sigma_{geo}(1 + \theta)
        \label{eq:eff}
    \end{equation}

    where $\sigma_{eff}$ is the effective cross section, boosted by gravitational focusing, $\sigma_{geo} = \pi R_1^2$ is the geometric cross section, $V_i$ is the incident velocity of the particles, and $V_e = \sqrt{\frac{2 G M_1}{R_1}}$, the escape velocity of the larger body. Here we've re-parameterized in terms of the `Safronov Number', $\theta = \frac{V_e^2}{V_i^2}$ \citep{Safronov_1972}. This approximation has been shown to be accurate over a wide variety of $\theta$ \citep{Wetherill_1985}.

    We consider a more general case, in which the primary body interacts with a secondary body of mass $M_2$, radius $R_2$, such that the two bodies orbit in circular orbits about their common center of mass with some semi-major axis $A$. These masses are assumed to be spherical and hence 
    \begin{equation}
        M_i =  \fracbrac{4 \pi}{3}  R_i^3 P_i
    \end{equation} 
    where $P_i$ are the densities.  As in the single body case, we assume these objects orbit within an accreting stream of test particles that move orthogonally to the center of mass with some velocity $V$ relative to that center of mass. 
    Finally, we assume the total mass flux due to accretion is small compared to the mass of the bodies, and hence the motion of the incoming particles can be approximated by test particles.  This also implies that significant changes in mass or angular momentum of the bodies occur slowly over many orbits. It is convenient to rescale our variables, we'll rescale our masses by $M_1$ and our lengths by $R_1$ and our velocities by $V_e$. This allows the problem to be described in terms of just a handful of free parameters. We denote our rescaled quantities with lowercase letters. We have $a = \frac{A}{R_1}$, $r_2 = \frac{R_2}{R_1} = \fracbrac{M_2 P_1}{M_1 P_2}^{1/3} = \left(m_2 \rho \right)^{1/3}$, $m_2 = \frac{M_2}{M_1}$, and $v=\frac{V_i}{V_e}$ where we've defined $\rho = \frac{P_1}{P_2}$ as the ratio of the densities of the primary and the secondary. Finally, we can assume without loss of generality that $M_1 > M_2$ and then set $M_1 = 1$ and $R_1 = 1$. Finally, we assume the two objects are of similar size and compositions and thus have similar densities which implies $\rho = 1$. The problem has now been reduced to a scattering problem with three rescaled, free variables, $m_2$, $a$, and $v$. 

    In the traditional one-body case, the problem is radially symmetric and has a closed-form solution. The two-body case is more complicated as it loses radial symmetry and since no general closed form can be obtained for the dynamics.  
    \zm{Additionally, the two body dynamics also has an additional degree of freedom due to the relative inclination between the orbital plane of the binary and the incoming stream of bodies. In general, for accretion for a disk, this will be controlled by the relative vertical and radial velocity dispersions in the disk. The ratio of these velocity dispersions is generally less than one, $\approx 0.5$, so we would expect radial flow to dominate. In general the vertical dispersion is smaller than the radial by about a factor of two, so encounters will typically follow the bulk motion of the disk and will tend to be aligned with the orbital plane \citep{IDA_1992,IDA_1993,Kokubo_2012}.  Though, for practical use special care will need to be taken to ensure this regime is satisfied for a given body undergoing accretion.  We might expect cross sections at other inclinations to increase the cross section slightly, since only cases near alignment will cause mutual shadowing of bodies from the accreting stream of mass.}
    Furthermore, unlike in the one-body case, not all trajectories are guaranteed to either impact a body or become distant from it on a short timescale.  Therefore, in general, the binary problem has three effective cross sections.  They are $\sigma_{1,eff}$ and $\sigma_{2,eff}$ which correspond to the cross sections of the primary and secondary bodies and $\sigma_{c,eff}$ which is an effective cross section of co-orbital configurations, \zm{describing those particles which do not exit the box during the simulation}. Given the complexity of the problem, a numerical approach is necessary. 

    We approach this problem by numerically integrating trajectories of particles with \texttt{REBOUND} for a given initial satellite true anomaly $\phi$ \citep{REBOUND} \footnote{https://github.com/hannorein/rebound}.  In our simulations, we take the binary to orbit in the xy plane.  We generate a grid of test particles at distance $d_x = 10 b_y$ away from the center of mass of our binary along an arbitrarily chosen x-axis. We define the maximum impact parameter along the y-axis as $b_y$. It is critical that we don't pick $b_y$ to be so small as to underestimate the cross section.    In the single body case, the maximum impact parameter that results in a collision $b$, is given by $b = R_1 \sqrt{1+\theta}$. Hence, the equivalent naive estimate for two bodies assuming all the mass in the system lies at a fictional body at the binary center of mass would be $b \approx R_{12}\sqrt{1+\theta_{12}}$ where $R_{12} = 1 + \fracbrac{m_2}{m_1}^{1/3}$ is the radius of the center-of-mass body and $\theta_{12}$ is its Safronov number. We can take into account the separation of the two bodies by simply adding a buffer to the cross section \zm{of size $a(1-m_2/2)$ to account for the distance of the bodies from the barycenter}. Finally, since our approach is only approximate, we increase the bound by a factor of $3$, which we found was sufficient to sample the entire cross section.  Hence, we take 
    \begin{equation}
        b_y = 3 R_{12} \sqrt{1+\theta_{12}} + a \left( 1 - \frac{m_2}{2} \right).
        \label{eq:bymax}
    \end{equation}

    Since the orbit is in the xy-plane, the binary extends less far along the z-axis, and hence the maximum impact parameter along the z-axis is smaller
    \begin{equation}
        b_z = \frac{3}{2} R_{12} \sqrt{1+\theta_{12}}.
        \label{eq:bzmax}
    \end{equation} 
    
\begin{figure}[!ht]
    \centering
        \includegraphics[totalheight=8.5cm]{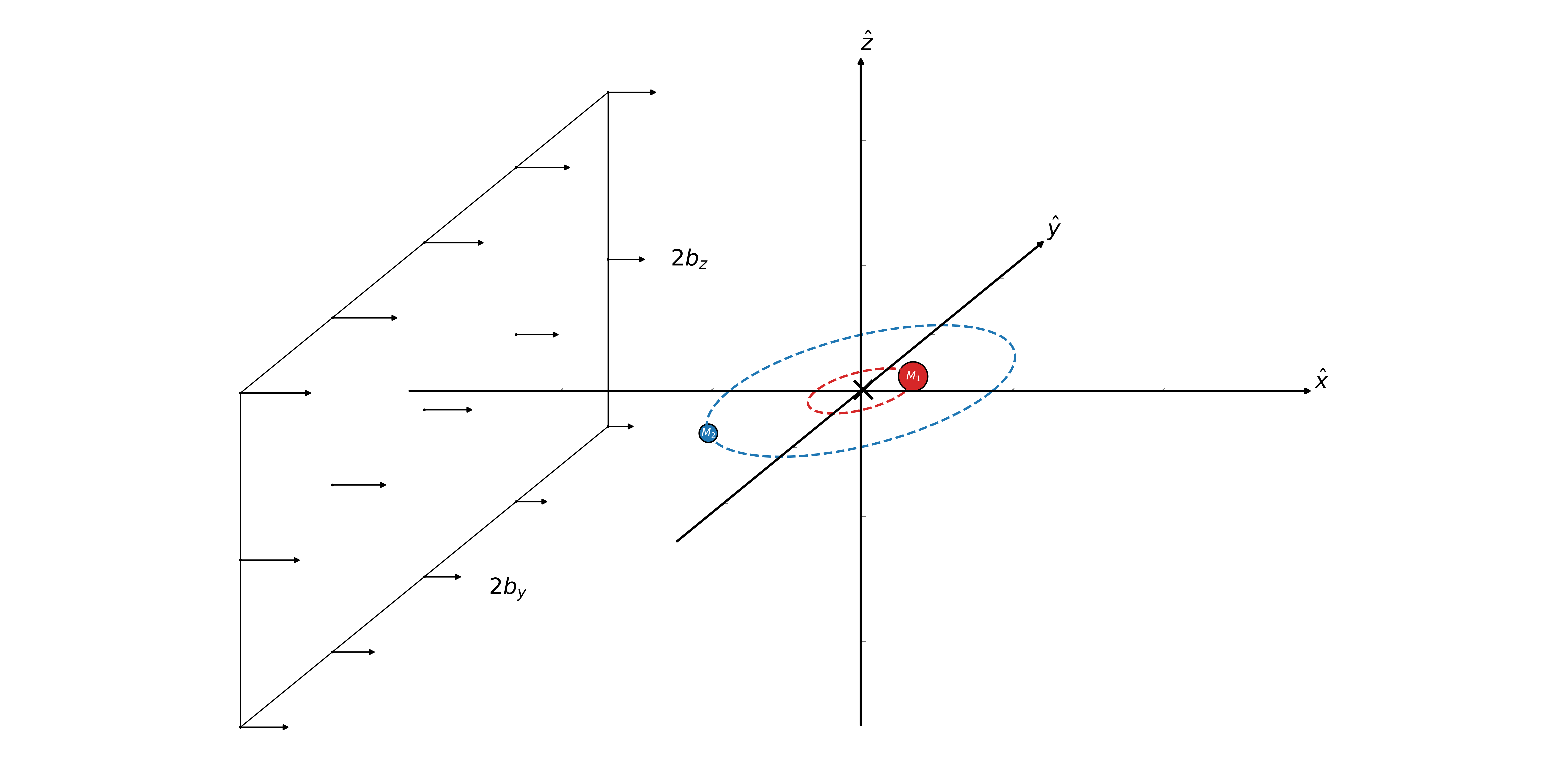}
        \caption{A mock-up of the numerical experiment. The binary pair orbits in the xy plane, with an incident grid of particles approaching it along the x axis. There are two separate impact parameters $b_z$ and $b_y$ which parameterize the size of the incoming grid of particles, whose values are a function of the binary's properties.} 
        \label{system_mockup}
    \end{figure}

    We then generate a grid of $N$ test particles, equally spaced in y from $-b_y$ to $b_y$ and in z from $-b_z$ to $b_z$, and integrate these to determine our cross sections.   We integrate our particles in a box, in which any particles that travel more than $10 b_y$ away from the binary are removed, a diagram of the initial condition can be found in Fig(\ref{system_mockup}).  We conduct our integrations until $t_{max}=40 b_y /v$, this is twice the time it takes for a typical incident particle to cross the simulation. Since the two-particle case is not symmetric with respect to the direction of the incoming particles (since the angle $\phi$ between the incoming particles and the binary can vary) we compute an average cross section by conducting this integration over $50$ points in $\phi$ ranging from $0$ to $2 \pi$ and average them.  Therefore each cross section we compute is the result of averaging $5 \cdot 10^7$ individual \zm{particle trajectories}. \zm{These numerically computed cross sections, which we call $\sigma_{1,num}$,$\sigma_{2,num}$ and $\sigma_{c,num}$ are computed by how many particles impact the primary body, how many impact the secondary body and how many remain in the simulation box at the end of the simulation. }

\begin{figure}[!ht]
    \centering
        \includegraphics[totalheight=12.5cm]{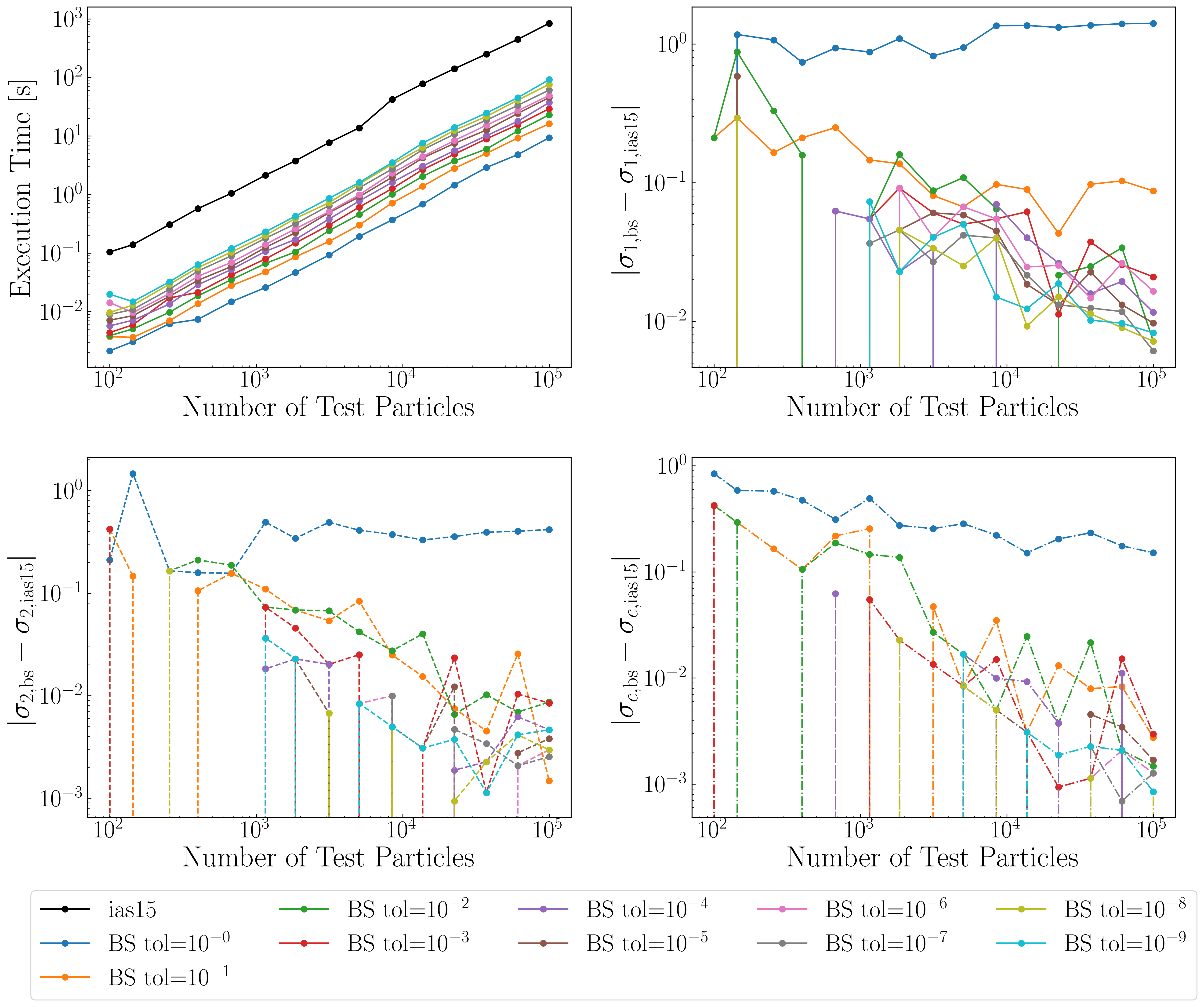}
        \caption{In the left panel we show the execution times of our various integrator configurations as a function of the number of test particles.  All of the tested integrators show power-law behavior.  In the right panel, we show the absolute difference between the primary cross sections derived with our \texttt{BS} integrator and that found with \texttt{ias15}.  We see that if the tolerance is set lower than $10^{-3}$ each integrator converges to the \texttt{ias15} result.  Lower tolerances result in convergence - but to the wrong value.  Vertical lines occur when the \texttt{ias15} and \texttt{BS} integrators compute the exact same cross section.  This happens rarely, but is more frequent in simulations with fewer test particles, as they are coarser.} 
        \label{optimization}
    \end{figure}

    Our simulations require an adaptive integrator since computing cross sections involves close approaches between particles. At the time of writing, \texttt{REBOUND} has two such integrators: there is the 15th order \texttt{ias15} integrator, best for high accuracy integrations, and a Gragg-Bulirsch-Stoer (\texttt{BS}) integrator, most suitable for medium accuracy integrations \citep{Everhart_1985,Hairer_1993,Rein_2015}. To compare these integrators we ran a trial integration of the system with varying numbers of test particles with both the \texttt{ias15} integrator and the \texttt{BS} integrator with varying error tolerances. For each tolerance, we computed the cross section of the primary and checked the absolute difference between the derived cross section and that found by \texttt{ias15}.  These results, in addition to the integration times, are shown in Fig(\ref{optimization}).   Overall, we find that for our purposes, setting the \texttt{BS} tolerance to $10^{-3}$ or better results in cross sections that converge to the \texttt{ias15} value. Tolerances greater than this \zm{threshold} resulted in convergence to incorrect cross sections and tolerances that were smaller all converged similarly (cross sections for the secondary and co-orbital states were also checked, and show similar behavior).  In terms of execution time, the \texttt{BS} integrators all perform well, being as much as two orders of magnitude faster than \texttt{ias15}.  Taking these results into account we elected to conduct our simulations with the \texttt{BS} integrator with a tolerance of $10^{-5}$, and a number of particles. This is two orders of magnitude smaller than what is needed for convergence - yet still runs an order of magnitude faster than the \texttt{ias15} integrations. In terms of precision, the error on the derived cross section scales roughly as the square root of the number of test particles - but much more gradually with the accuracy of the integrator.  Hence, to obtain a precise cross section it's better to perform slightly less accurate simulations with many particles over highly accurate ones with few.  We choose $N=10^6$ particles for our sets of simulations. 
    
    After optimizing our integrator, we must optimize our collision detection algorithm.  While \texttt{REBOUND} includes a collision algorithm by default, it checks all particles at every step for collisions and is thus $\mathcal{O}(N^2)$, with $N$ the number of particles being integrated. However, since in our case, we care only about collisions with either the primary or the secondary - not between test particles - this approach is inefficient. Therefore, we modify the default \texttt{REBOUND} algorithm to exclude test particles from its collision checks, rendering the algorithm $\mathcal{O}(N)$, where $N$  is the number of test particles.  This dramatically speeds up the integration. 
     
\begin{figure}[!ht]
    \centering
        \includegraphics[totalheight=8.5cm]{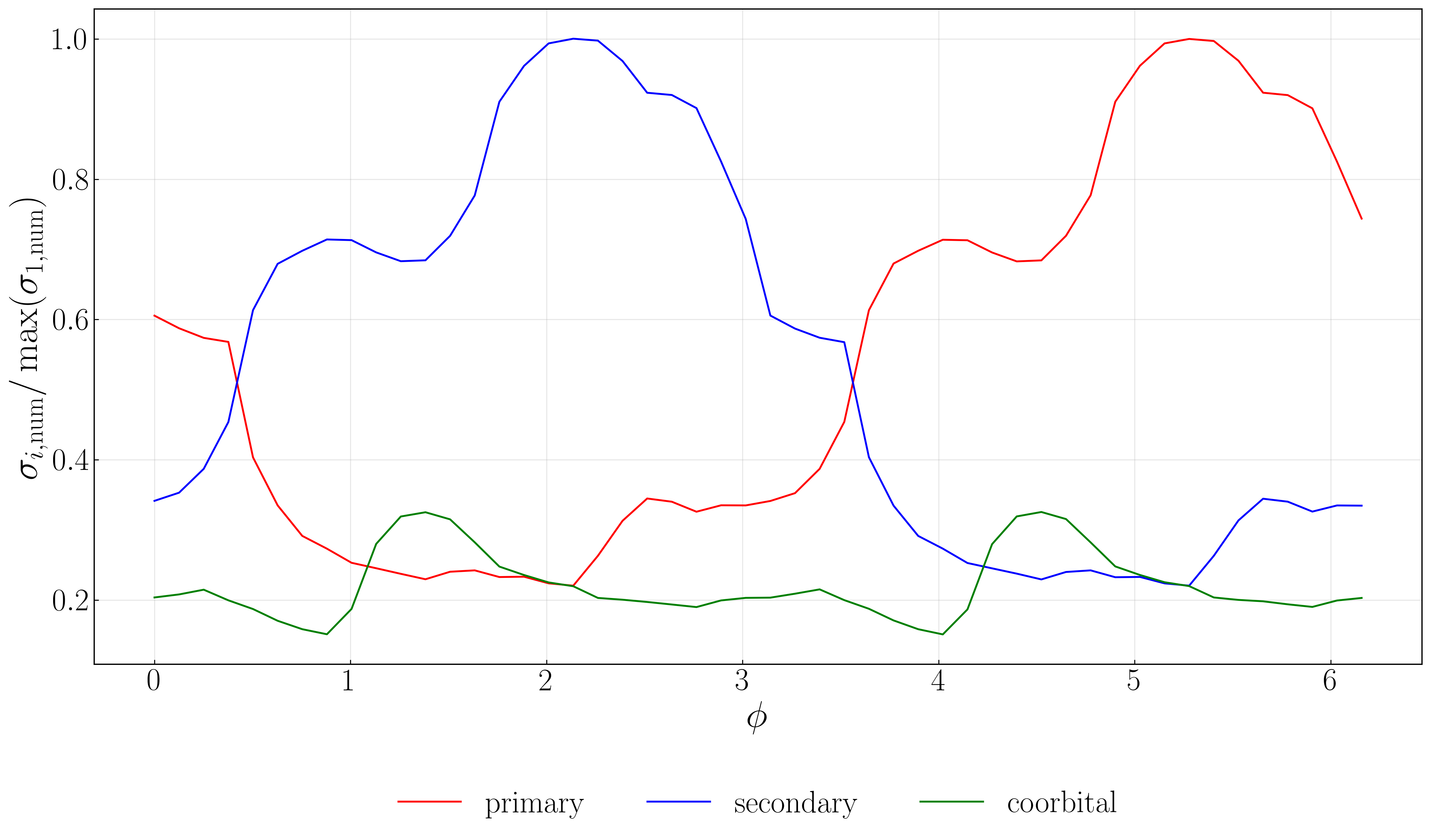}
        \caption{We show the normalized cross sections of a the binary as a function of the initial relative angle ($\phi$) between the secondary and the $x$ axis.  The cross section can changes considerably as a function of this initial angle.} 
        \label{crossvstheta}
    \end{figure}

\begin{figure}[!ht]
    \centering
        \includegraphics[totalheight=10.5cm]{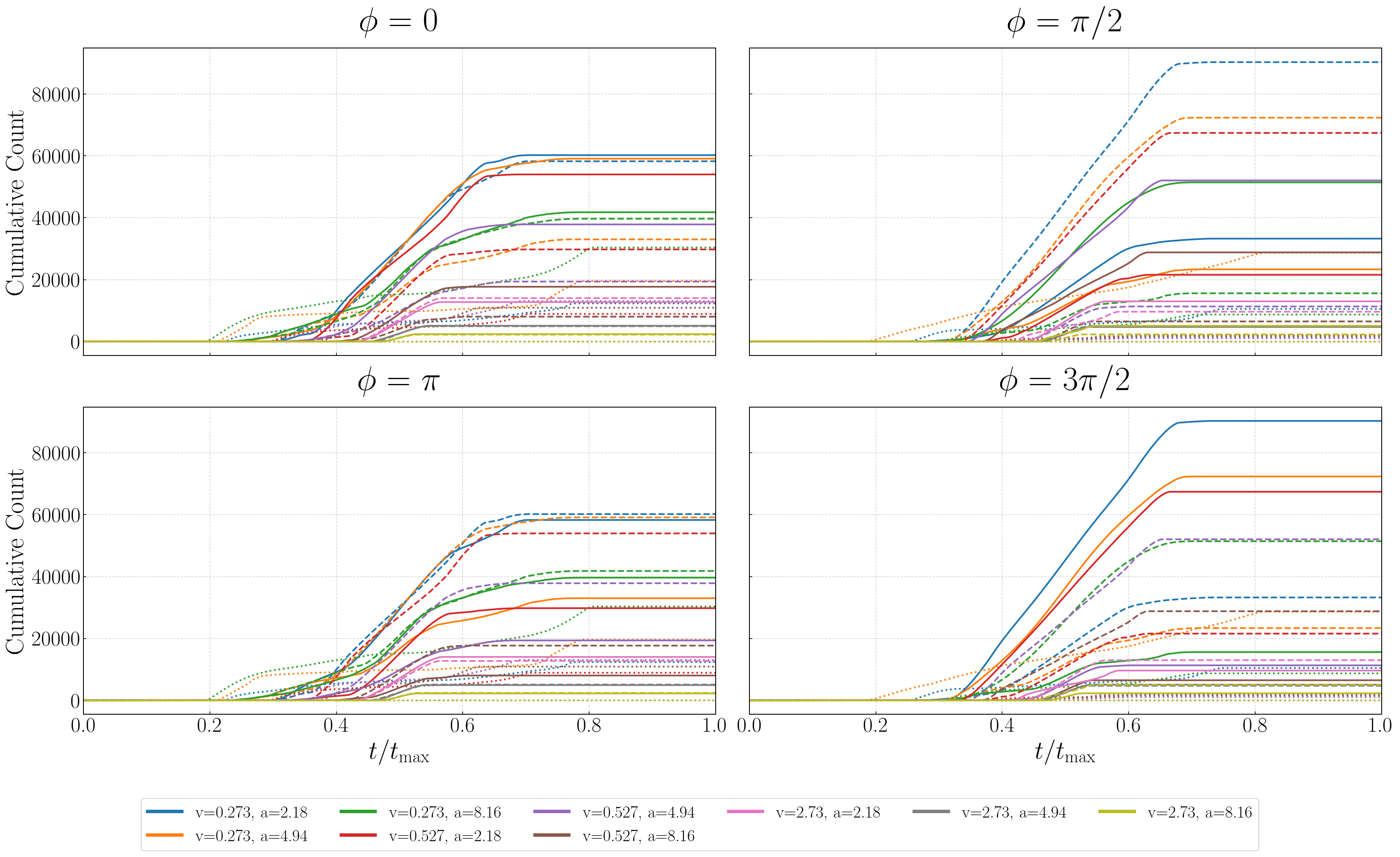}
        \caption{The cumulative number of particles impacting the primary (solid lines), secondary (dashed lines) or exiting the simulation without hitting either body (dotted lines) as a function of time, for a variety of $v$ and $a$. In all cases shown here $m_2=1$. We see in general our cross sections converge quickly over the simulation and typically do not change after $t/t_{max} \approx 0.8$} 
        \label{crossvstime}
    \end{figure}

    Since we obtained $b_y$ and $b_z$ by rough approximation, we check to show that this choice is sufficient numerically \zm{on every simulation we run}, by ensuring that the collection of points with the largest impact parameters - which form the borders of the generated grid of points - do not interact with either body over the duration of the simulation. \change{The convergence of these simulations over time, as well as a function of the initial angle $\phi$ is shown in Fig \ref{crossvstheta} and Fig \ref{crossvstheta}.  Fig \ref{crossvstheta} shows the expected periodicity in phi at the resolution of our grid (50 points in theta) showing the computed cross section can change significantly as a function of $\phi$, and so our averaging procedure is necessary.  Fig \ref{crossvstime} shows the convergence of the cross sections over time, suggesting our cross sections are converged by the time we conclude or simulations.  While we only show a few cases, the plots are similar for all the $v$ and $a$ we check.}

    Finally, we must determine reasonable bounds on the grid of $m_2$, $v$, and $a$ over which we carry out our simulations. We consider a grid with $20$ by $20$ points with linearly spaced $m_2$ and $a$ such that $0 < m_2 < 1$ and $1.26 < a <10$. We choose a lower bound of $1.26$ on the $a$ as this is the Roche limit of two bodies with the same density.  We compute this grid of points for $15$ different logarithmically spaced $v$ such that $0.1 < v < 10$, chosen to sample a wide variety of focusing behaviors.  Finally, we exclude any points in the grid where the two planets touch, guaranteeing that $1 + r_2  > a$. These choices require the computation of $5535$ cross sections.  The results of this paper are, therefore, a grid of points that can be interpolated over for future studies and are the result of over a quarter trillion individual simulations. 

\section{Results: Limiting Cases}

    While in general estimating the cross sections must be done numerically, since trajectories in the three-body problem, even in the circular restricted case, have no closed-form solution, certain limits of the problem do permit closed-form approximations that can be compared to our results as verification. Several limits are worth explicit consideration \zm{and serve to verify our approach}.

        \subsection{The high $v$ limit}
        \label{sec:limit1}

        In the case where the incident velocity is very large compared to the escape velocity (small Safronov limit) gravitational focusing is negligible.  Hence, \zm{we can treat the situation approximately as a} two-body limit in which the trajectories are completely unperturbed by the gravity of the massive objects - implying the test particles move on straight lines.  The \zm{accretion of the primary and secondary, in this limit are independent}, except for those \zm{geometries} in which they shield each other from the incoming stream of particles.   \zm{The mathematical description of this case, therefore, is nearly identical to that seen in the transit of a planet in front of a uniform radiator - as in the physical situation is identical, except that the test particles in the transiting case are the photons from a star. Therefore we borrow heavily from the transiting planets literature to derive the approximate cross sections in this limit.}
    
        While the fraction of the area seen in a mutually eclipsing system of two bodies does have a closed, analytic form \citep[e.g.][among others]{Kipping_2010,Luger_2019},  its form is complex and unwieldy. Consequently, it is useful to make approximations that capture the most dominant behavior. \zm{Since there is no gravitational focusing in the high velocity limit,} the average area seen by incoming particles with high velocity over a given period is just the cross section of the body in question
    
        \begin{equation}
            \sigma_{eff,i} = \frac{1}{T} \int_0^T \sigma_{geo,i} (1 - \delta(t)) dt.
        \end{equation}
    
        Where $T$ is the orbital period of the two bodies and  $\delta$ is the relative reduction in the cross sectional area seen by incoming particles when one body eclipses another. We model the relative eclipses as trapezoidal transits, the total duration of a mutual eclipse (regardless of which body eclipses which) can be written as 
        \begin{equation}
            \tau_t = \frac{T}{\pi} \arcsin\fracbrac{R_1 + R_2}{A}
        \end{equation}    
    
        whereas the duration over which the transit is flat is given by  
        \begin{equation}
            \tau_f = \frac{T}{\pi} \arcsin\fracbrac{R_1 - R_2}{A}
        \end{equation}    
    
        both expressions were adapted from \citet{Seager_2003}.  When the transit occurs, the relative cross sectional area of the primary will be reduced by $\delta = \fracbrac{R_2}{R_1}^2$. The area of the secondary, in contrast, will drop to zero as it is completely occulted by the larger primary, so $\delta = 1$.  If we treat the transit as a trapezoid, with linear ingress and egress, this average takes a particularly simple form as, 
    
        \begin{equation}
            \sigma_{eff,i} \approx \frac{\sigma_{geo,i}}{T} \left( T - \frac{1}{2} \delta(\tau_f + \tau_t) \right).
        \end{equation}
    
        It is now useful to invoke the approximation $\arcsin(x) \approx x$.  While the approximation is closest for small x, the relative error $\frac{x-\arcsin(x)}{\arcsin(x)}$ is less than $10 \%$ so long as $x \lesssim 0.7$.   With this approximation, the enhancement factors of the two bodies can be written, remarkably simply, as
        
        \begin{eqnarray}
            \frac{\sigma_{eff,1}}{\sigma_{geo,1}} = 1 - \frac{r_2}{a \pi} \\        
            \frac{\sigma_{eff,2}}{\sigma_{geo,2}} = 1 - \frac{1}{a \pi}.
            \label{eq:approx:fast}
        \end{eqnarray}
    
\begin{figure}[!ht]
        \centering
            \includegraphics[totalheight=6.5cm]{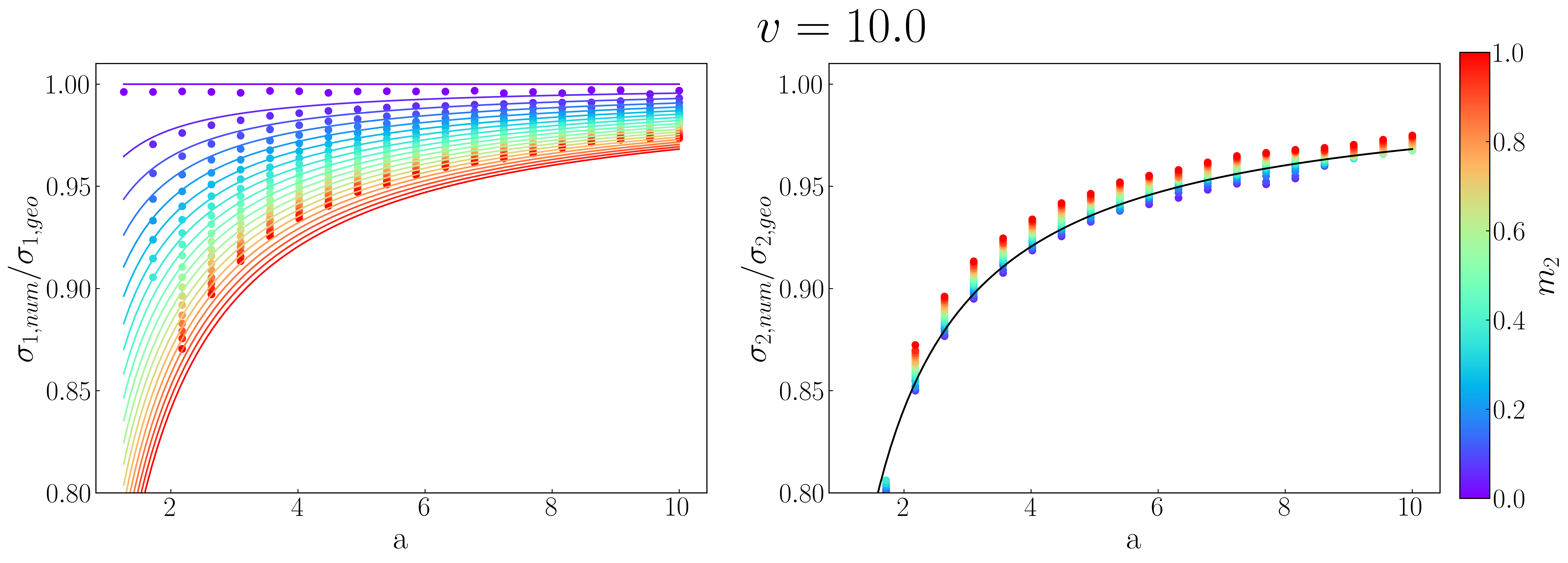}
            \caption{ \zm{Here we show the computed numerical cross sections scaled by the geometric cross section $\sigma_{num}/\sigma_{geo}$ for each body in the system for several mass ratios against the predicted cross sections in the analytic limits derived in Eq(\ref{eq:approx:fast}). \change{The analytic limits are solid lines and the numerically computed values are the scatter points.}  The first panel shows the relative cross section for the two bodies secondary against  for very high velocities.  We show the results for the the high velocity limit. In this limit, gravitational focusing is small, and shielding results in the effective cross sections of both bodies being smaller than their geometric cross sections. While the gravitational focusing factor is small, \change{and so our assumption of no focusing in Eq(\ref{eq:approx:fast}) is appropriate, the approximation} shows a small systematic vertical offset between the analytic and numerical predictions.  This leads to the overestimation of the cross section in the limit $m_2 = 0$} }
            \label{highvcase}
        \end{figure}

        \zm{While approximate, Fig(\ref{highvcase}) shows our approximate expressions describe the results of the simulations rather well. In the left panel we can see the cross section of the primary decreases with the increasing mass (and thus radius) of the secondary. Intuitively, this is simply because a larger body blocks more of the incoming particles than a smaller one.  In the right hand panel we show that the cross sections for the secondary is effectively constant with the mass of $m_2$, this follows from the symmetry in Eq(\ref{eq:approx:fast}), as the cross section depends on the radius of the primary which is fixed to $1$ in our simulations. }
    
        \zm{Finally, we note that a limiting case similar to the high $v$ limit exists when the bodies are sufficiently widely spaced that they move negligibly over the course of the simulation, but still permit significant gravitational focusing.  In this case, the dynamics of their motion can be neglected and the system will be well approximated by the Euler-three-body problem, detailed in \citet{Biscani_2016}.   While this problem has a closed-form solution, it is unwieldy and hence cross sections in this limit are best computed numerically.}

    \subsection{The equal mass limit}
    \label{sec:limit2}
    
\begin{figure}[!ht]
    \centering
        \includegraphics[totalheight=6.5cm]{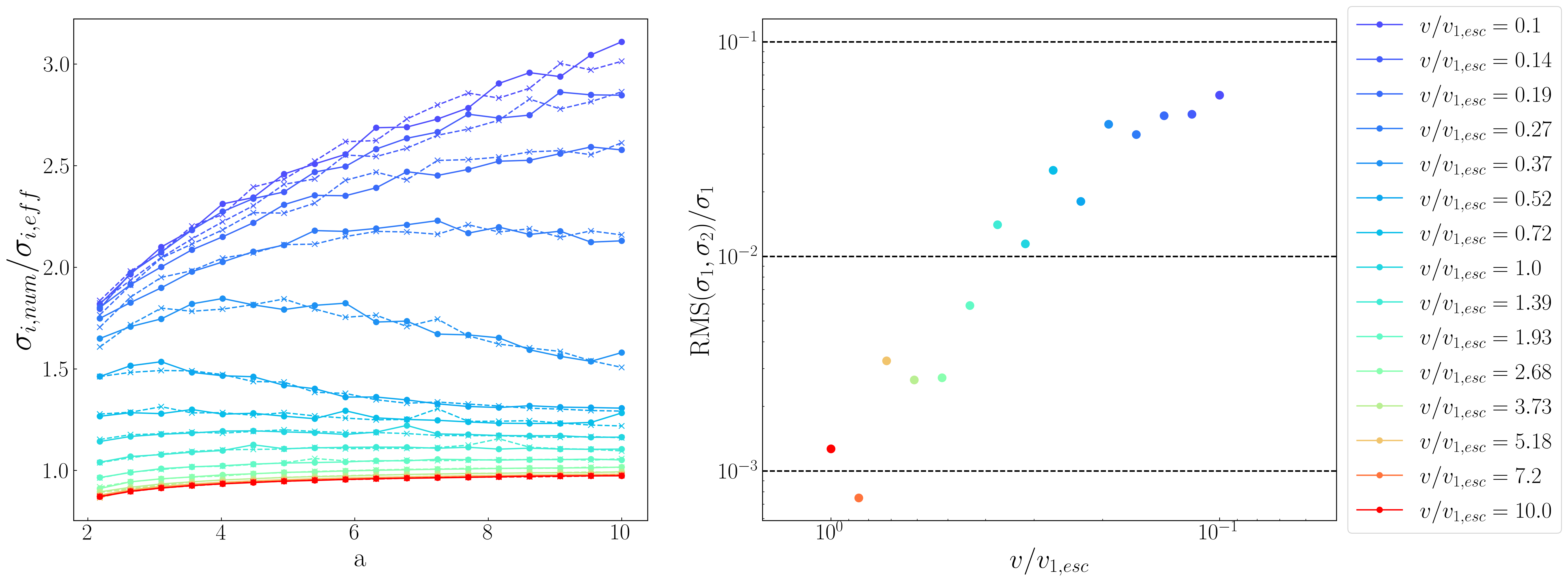}
        \caption{In the left panel we show the computed numerical cross sections scaled by the effective cross section derived in Eq \ref{eq:eff} for both the primary (solid lines) and the secondary (dashed lines) for a variety of different $v$. In each case, the cross sections of the two bodies closely follow each other, as would be expected by symmetry.  In the right-hand panel, we show the root mean square difference in the cross section relative to the effective cross section ($\mathrm{RMS}(\sigma_1, \sigma_2) =\sqrt{\langle (\sigma_1 - \sigma_2)^2 \rangle}$). While our errors increase with decreasing $v$, they remain below $6 \%$ for all the simulations} 
        \label{fig:equalmass}
    \end{figure}

    Another useful limit occurs if $M_1=M_2$.  In this case, the simulation is fully symmetric and there should be no difference between the cross sections of the two bodies.  This observation yields no additional physical insight. However, it can serve as an additional check on our simulations and is therefore worth exploring. We show a comparison between our derived cross sections of the primary and secondary bodies in Fig(\ref{fig:equalmass}).  \zm{Since the system is symmetric in this limit, any deviation from perfect symmetry we observe in our cross sections must be due to where exactly we initialize our test particles and on our $\phi$ grid.  Hence checking the error in the equal mass limit provides us a notion of how the coarseness of both grids effect our results}. These changes in the grid coarseness are partly due to changes in the $a$ which changes by a factor of $5$ over our simulations but are dominated by changes in the gravitational focusing factor $\sqrt{1+\theta_{12}}$, which changes by two orders of magnitude (the grid in $\phi$ is always constant).  It is therefore reasonable to approximate these errors as being only a function of $v$. To estimate the error we consider the root mean square error between the primary and secondary when $m_2 = 1$ for all computed cross sections with a given velocity $v$. This reveals, as shown in Fig(\ref{fig:equalmass}), that the error in general increases towards smaller $v$ and larger gravitational focusing factors.  This is expected, as in these cases, the $10^6$ test particles are spread over a larger area and therefore have lower effective resolution. However, even in the most extreme cases,  the grid coarseness results in no more than a $6 \%$ error in the computed relative cross section.  
   
\section{Results: General Cases}

\begin{figure}[!ht]
    \centering
        \includegraphics[totalheight=4.5cm]{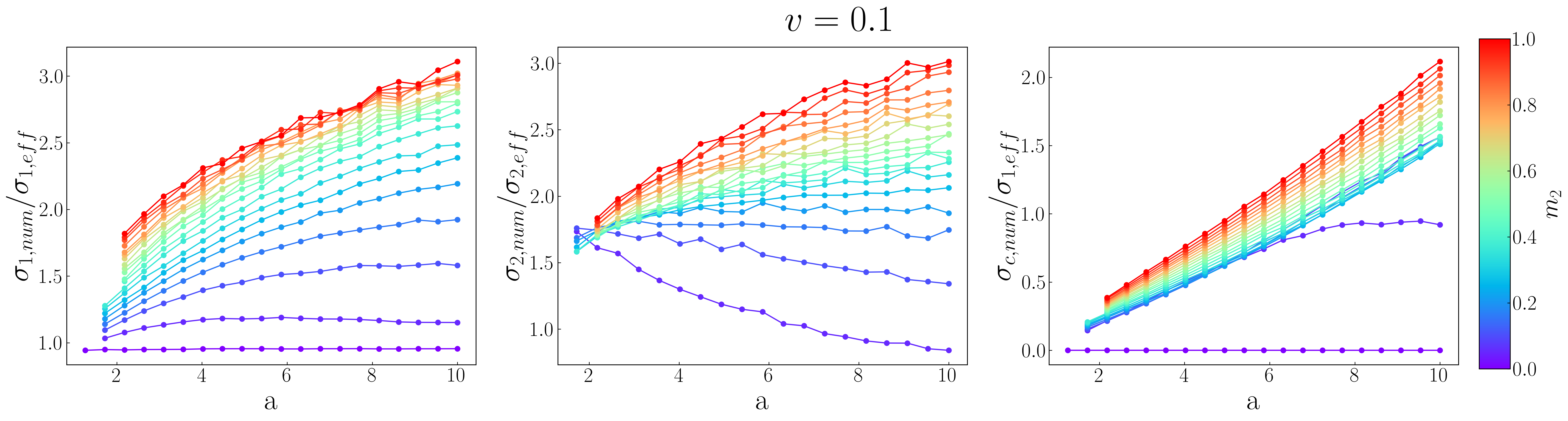}
        \includegraphics[totalheight=4.5cm]{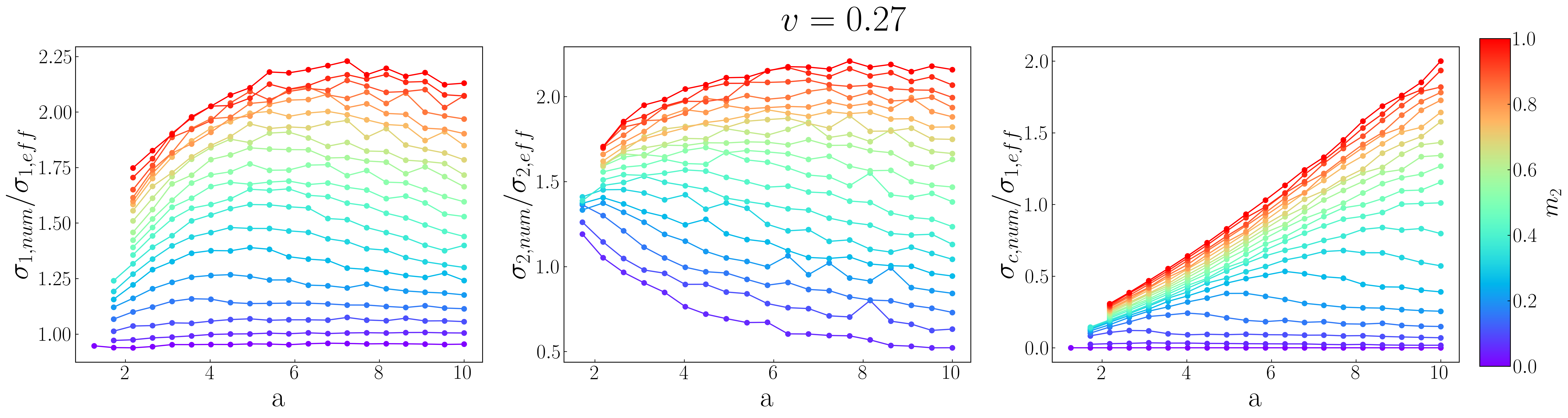}
        \includegraphics[totalheight=4.5cm]{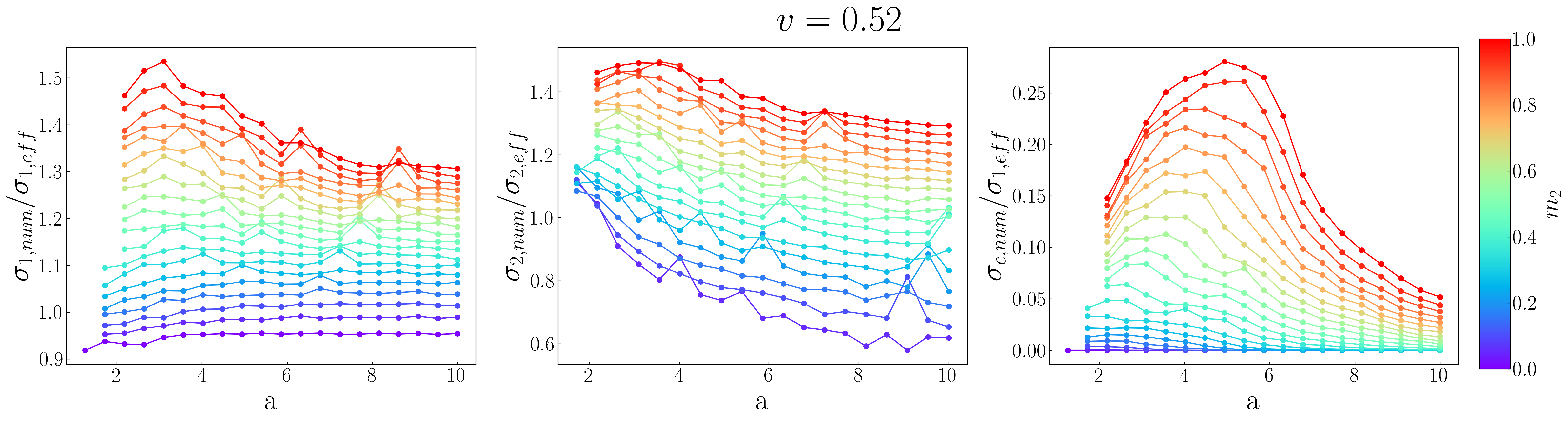}
        \includegraphics[totalheight=4.5cm]{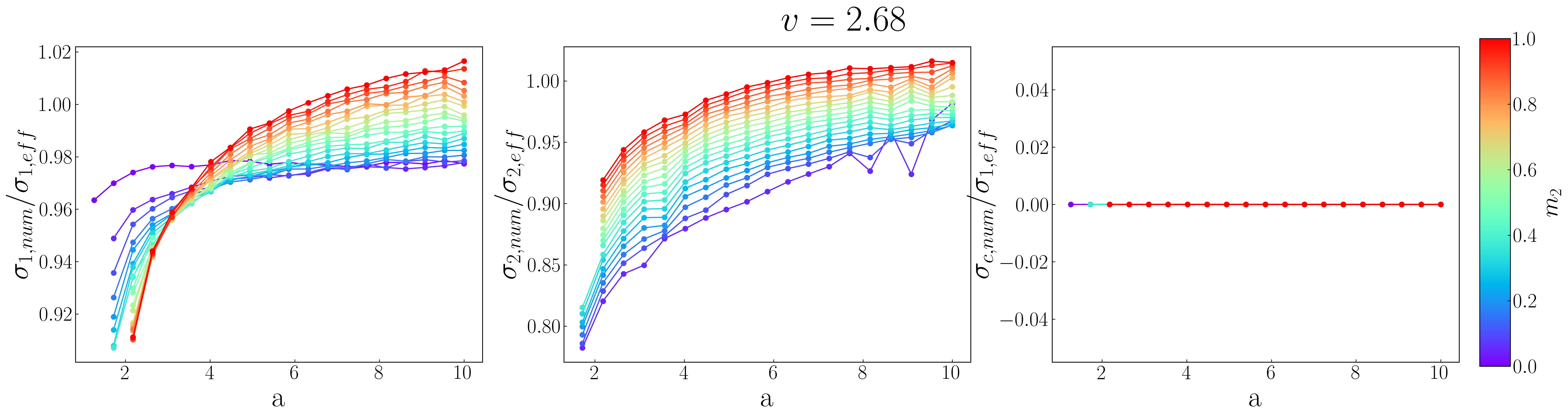}
        \includegraphics[totalheight=4.5cm]{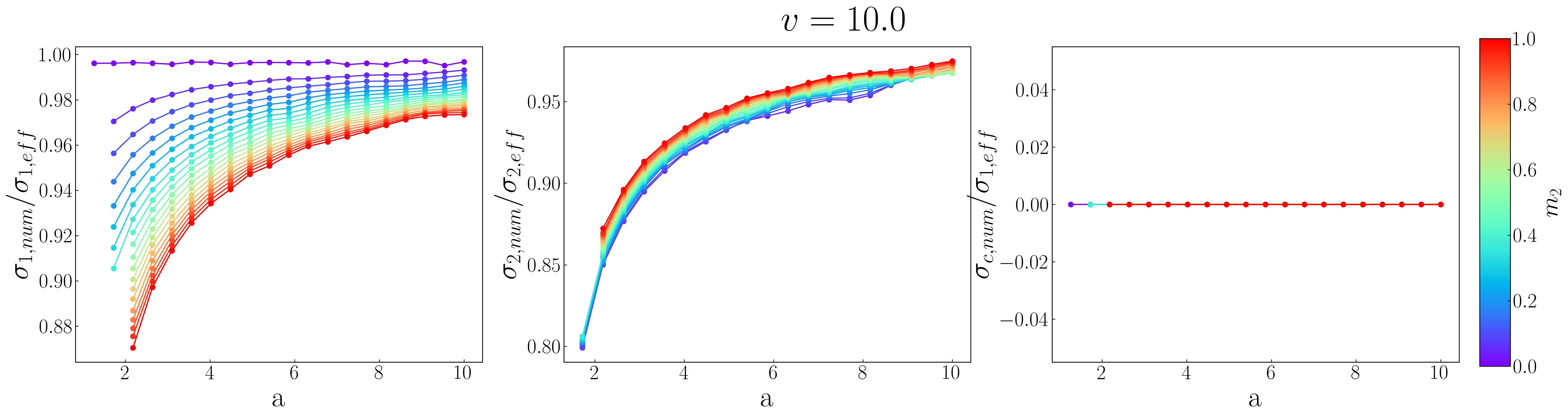}
        \caption{Here we show the numerically computed relative cross sections $\sigma_{num}/\sigma_{eff}$ for our system for several velocities and mass ratios. The first column shows the relative cross section for the primary, the second shows the cross sections of the secondary and the final shows the effective `coorbital cross section' normalized to be relative to the primary. While each row of plots is shown at the same scale, note the axes range changes between rows.  } 
        \label{cross}
    \end{figure}

    In general, the computed cross sections adopt a variety of behaviors as $v$, $m_2$ and $a$ are varied; these results are summarized in Fig(\ref{cross}).  Most notably, the effect of three-body dynamics is drastic, with the cross section being enhanced by as much as a factor of 3, or depressed by as much as a factor of 2, depending on the values of $v$, $a$ and $m_2$ being considered.  As described previously, in the high-velocity limit co-orbital states are absent and every particle either impacts a body or exits the simulation box. In this limit, mutual shielding of the bodies decreases their effective cross sections. As the velocity decreases ($v \approx 2.5$) and gravitational focusing increases, these decreases from shielding are increasingly canceled by the growing gravitational enhancement, most slowly in the case of the closest binaries.  These effects can be seen in the two high velocity cases ($v=10.0$ and $v=2.68$) in Fig(\ref{cross})
    
    At more moderate Safronov numbers, where $v \approx  0.5$, coorbital states begin to appear.  They occur most frequently at a critical, mass-dependent, $a$, which ranges from $2-6$ depending on $m_2$.  This behavior can be seen in the middle \change{row} of Fig(\ref{cross}).   In this regime, the co-orbital cross section is a small fraction of that of either body. The cross sections of the primary and secondary now peak at small $a$, in contrast to the high-velocity limit where larger separations yielded a larger cross section.  The cross sections of both bodies are generally higher than predicted from the single-body gravitational focusing approximation given in Eq(\ref{eq:eff}), however small secondaries, especially at large distances show a decrease in effective cross section by up to a factor of $2$ compared to the effective cross section they would have if isolated. 

    As the Safronov number increases further, with $v \approx 0.25$ the behavior changes yet again. Like the high-velocity limit, the effective cross section increases with increasing $a$, but instead of doing so monotonically, it peaks and then begins to decline.  This behavior can be seen in the second row of Fig(\ref{cross}). Here, at large $a$, we find the co-orbital cross section grows significantly and can become comparable to the cross sections of the two bodies.  This cross section grows approximately linearly with $a$, before turning over and decreasing, with the turnover point a function of $m_2$.   In this regime, the cross section of the primary body is always larger than expected by analytic theory, by as much as a factor of $2$ .  The cross section of the secondary ranges from being much larger, to much smaller than expected, depending on its $m_2$.   

    Finally, when the Safronov number is the largest, with $v \approx 0.1$, the behavior seen earlier becomes even more extreme.  The cross section of the primary becomes uniformly larger - up to three times what is predicted analytically, and continues to rise even at the maximum $a$ we simulate. It is likely that these cross sections also peak before falling, as when the two bodies are very widely spaced, with $a >> b_{1,max} + b_{2,max}$ the two bodies won't have much an effect on each other, and the one body case will be recovered.  The secondary also has a cross section greater than what would be predicted - except in the case of large $a$. This behavior can be see in the first row of Fig(\ref{cross}).
    
    \subsection{Initial Condition Portraits}

    Understanding the changing slope of the cross sections with $a$ as a function of $v$ seen in Fig\ref{cross} is of particular interest.  While each point in Fig(\ref{cross}) is the average of the results of $50$ different simulations over $\phi$, we can obtain some additional insight into some of the properties of the cross sections by examining examine the final states of the initial points for a given $a$, $v$, and initial angle $\phi$.  We limit ourselves to only the equal mass case for simplicity.  We create these portraits by plotting the initial conditions of the test particles color-coded by where they end up at the conclusion of the simulation.  Points that exit the simulation before it finishes are colored black, those that impact the primary body are colored red, those that impact the secondary are colored blue and those that do not impact either body or exit, are colored green. While the resolution of our simulations is adaptive, within each figure, each portrait has been scaled to the same resolution, so the cross sectional areas can be directly visually compared.  

    Since our system is equivalent to the well-studied circular restricted three-body problem, it's briefly worth considering the role of the Jacobi constant in these portraits. When expressed in cartesian coordinates in the rotating, center of mass frame, the Jacobi constant of a given particle can be written as

    \begin{equation}
        C_j = n^2 (x^2 + y^2) + 2 \left( \frac{G M_1}{D_1} + \frac{G M_2}{D_2} \right)
        - (V_x^2 + V_y^2)
    \end{equation}

    where $n$ is the mean motion of the binary, $D_i$ is the distance from the test particle to the $i$th mass, and $v_i$ is the $i$th component of the velocity in the rotating frame \citep[e.g][]{MDbook}.   In general, the circular restricted three-body problem permits three disjoint zero velocity surfaces, two around each body and a third outer surface surrounding them both.  A particle far from the binary can only ever impact one of the massive bodies if the outer surface intersects with one of the inner ones.  As the Jacobi constant decreases, this intersection typically occurs first at the $L_2$ Lagrange point (in the case of equal masses, it will occur at the $L_3$ and $L_2$ points simultaneously, since the system is symmetric). Hence the value of the Jacobi constant at $L_2$ is a critical value, above which no test particles can interact with massive bodies.  Since our particles only have an initial velocity in the $x$ direction, and the binary lies in the x-y plane, the Jacobi constant of the test particle grid is very nearly linearly proportional to the initial $y$ coordinate of the test particle.  Hence, the cutoff Jacobi constant takes the form of a vertical line, to the left of which test particles may impact the primaries, and to the right of which no such interactions are possible. The figures in this section we show this cutoff if it within the range of generated test particles.  Examination shows that the portraits for the two bodies are all consistent with the bound.
    
\begin{figure}[!ht]
    \centering
        \includegraphics[totalheight=12.5cm]{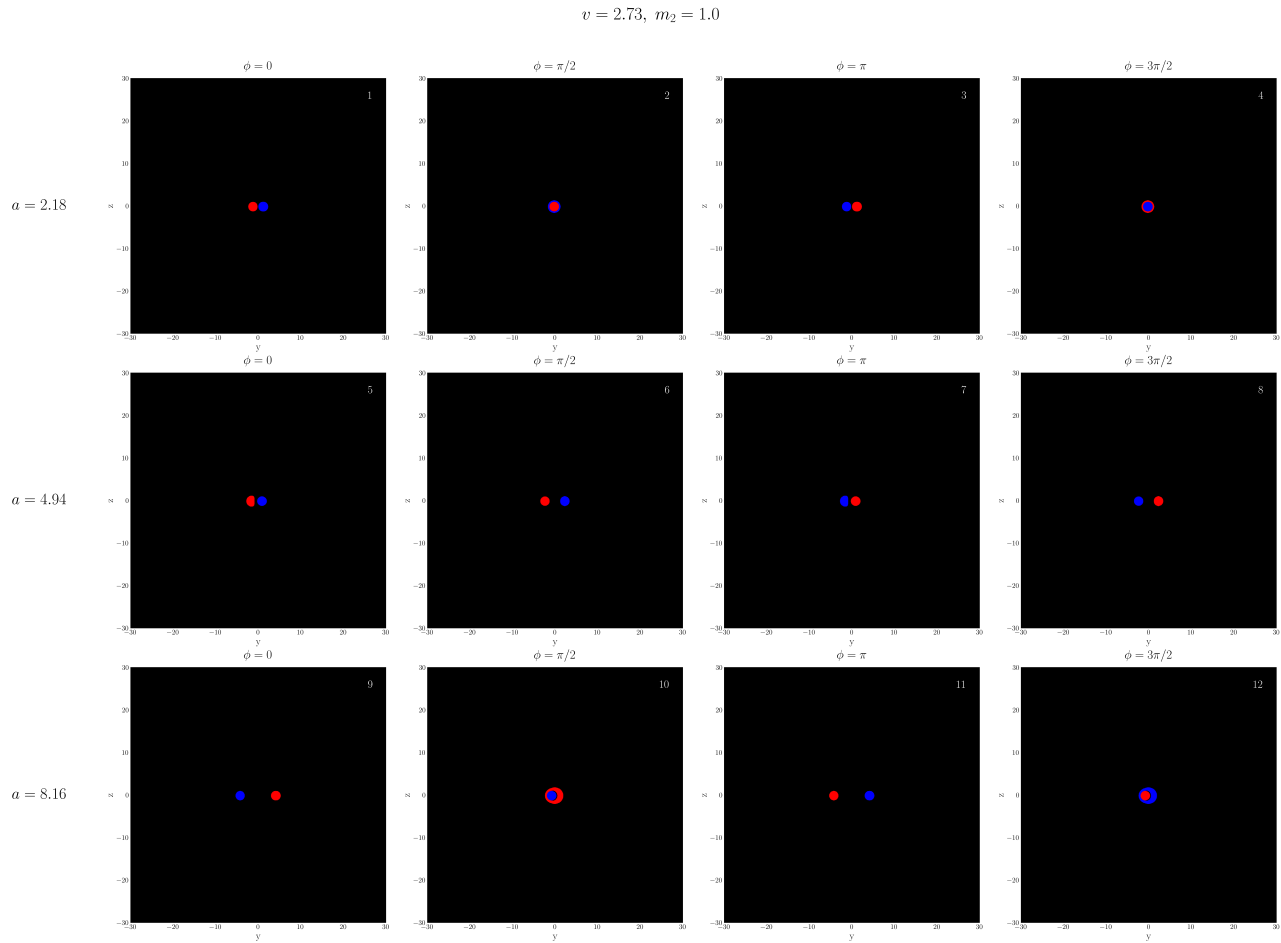}        
        \caption{Here we show the portraits for test particles of high incident velocity $v=2.73$, three semi-major axes, $a=2.18$, $a=4.94$, $a=8.16$ and four initial $\phi$. Points that exit the simulation before it finishes are colored black, those that impact the primary body are colored red, those that impact the secondary are colored blue and those that do not impact either body or exit, are colored green. All simulations are shown on the same scale.  } 
        \label{basins1}
    \end{figure}

    The cross sections are simplest when the initial velocity is high and gravitational focusing is minimal, shown in Fig(\ref{basins1}).  In this limit, there are a few separate behaviors worth noting. First, in the $a=2.18$ case the shielding behavior described in Sec (\ref{sec:limit1}) can be clearly seen in the $\phi=\pi/2$ and $\phi=3\pi/2$ portraits, where the two bodies shadow each other. Whereas in the $\phi=0$ and $\phi=\pi$ cases the test particles only interact with one body or the other unoccluded, resulting in circular, nearly geometric cross sections (subplots  1,3,5,6,8,9,11 in Fig(\ref{basins1}).  As the two bodies get further from each other, more nonlinear effects can be seen. Here the $\phi=\pi/2$ and $\phi=3\pi/2$ portraits are nearly unperturbed (note that the shift in the unperturbed case from $\phi=0$, $\phi=2\pi$ is due to the test particles first encountering the binary at a different phase in their orbits).  In the other two cases $\phi=0$ and $\phi=\pi$, one body retains a circular cross section where the other is distorted.  This distortion comes from the fact that the bodies are now significantly separated from each other in space, hence there is time for the gravity of one body - the one first encountered by the stream to affect the trajectories of particles before they can encounter the second body (subplots 2,3 in Fig(\ref{basins1}). In the last case, where the two bodies are widely separated, the shielding behavior becomes significantly more nonlinear, as there is even more distance over which perturbations from the first encountered body can build up into displacements. In this case, the portrait of the shielded body deviates significantly from a circle, even exhibiting a small hole where particles that would have otherwise would have impacted are deflected away (subplots 10,12 in Fig(\ref{basins1}). 

\begin{figure}[!ht]
    \centering
        \includegraphics[totalheight=12.5cm]{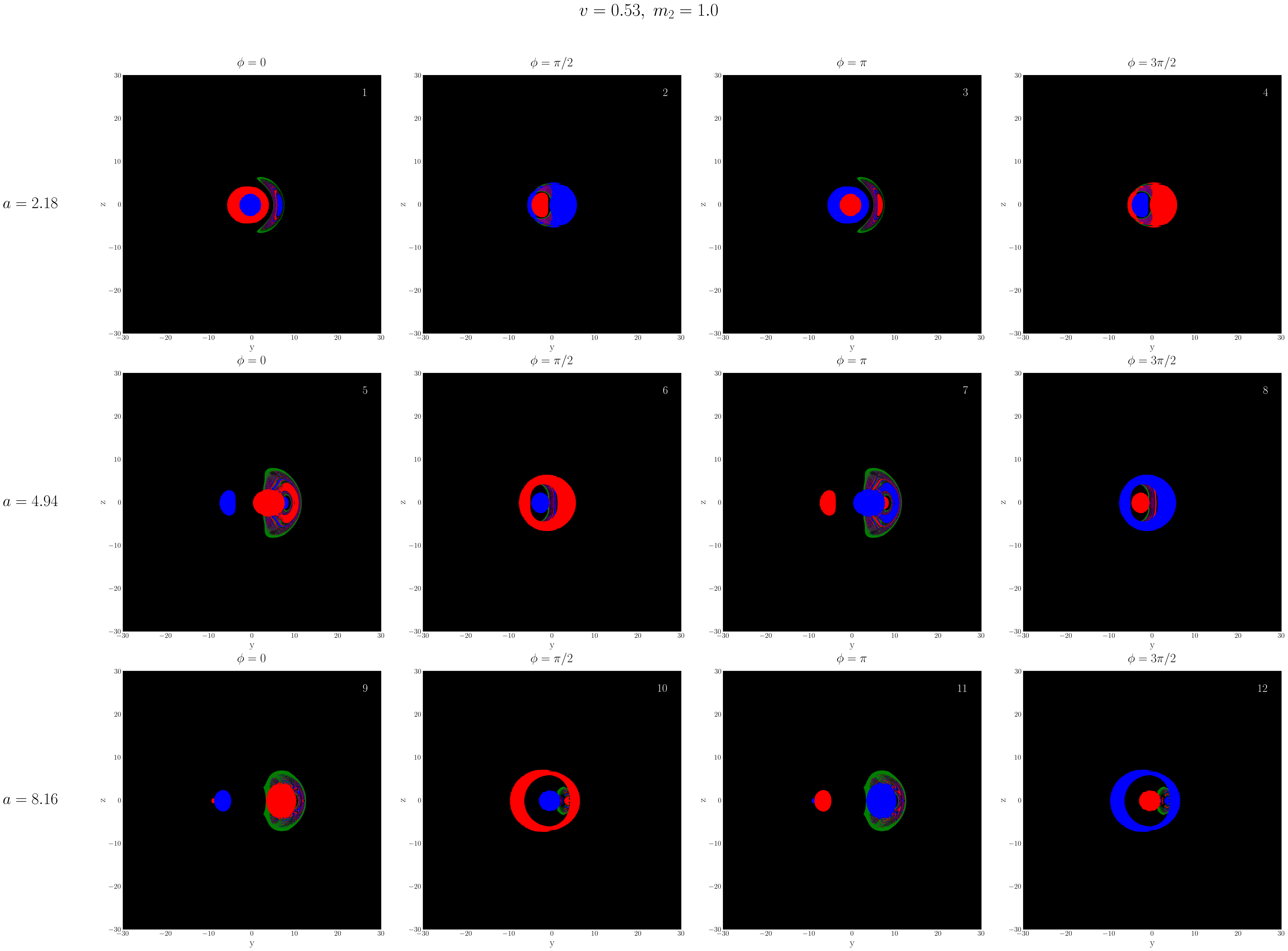}        
        \caption{Here we show the portraits for test particles of moderate incident velocity $v=0.53$, three semi-major axes, $a=2.18$, $a=4.94$, $a=8.16$ and four initial $\phi$. Points that exit the simulation before it finishes are colored black, those that impact the primary body are colored red, those that impact the secondary are colored blue and those that do not impact either body or exit, are colored green. All simulations are shown on the same scale} 
        \label{basins2}
    \end{figure}

    The portraits in the case of more moderate velocities, as shown in Fig(\ref{basins2}), are much more complicated, the simplest case, where $a=2.18$ can be understood to be a more extreme case of the $a=8.16$ dynamics in the high-velocity regime shown in Fig(\ref{basins1}). Here the distortion and shadowing effects of one body on another can be seen prominently since the lower velocity of the incoming particles increases the length of time gravity can act upon it (effectively this is the same effect as the wider spacing).  In addition, for certain $\phi$ we see the development of an island of encountering trajectories separate from the main portraits in the center of the plot that we saw in the high-velocity case (subplots 1,3 in Fig(\ref{basins2})).  The fates of incoming particles in this island are complex, with particles encountering the primary, secondary, or entering coorbital states. However, in the moderate velocity case, as the separation increases, the portraits show substantially different behavior than what was seen in the high-velocity case.  The islands we previously observed begin to change shape and grow (subplots 5,7,9,11 in Fig(\ref{basins2}). For certain $\phi$ we see the connection of the interacting island with the main portraits and the formation of a separate shallow ring of trajectories (subplots 6,8 in Fig(\ref{basins2}).  In other cases, we see large gaps open up in the portraits of the shadowed object. While the development of these gaps only occurs for certain $\phi$, the overall trend is to decrease the cross section.  The $a=8.16$ case takes this to an extreme, with large gaps in the shadowed object opening up, reducing its cross section to a thin ring (subplots 10,12 in Fig(\ref{basins2}).  This change in morphology of the initial condition portraits suggests an explanation for the decreasing cross section with $a$ for moderate $v$.

\begin{figure}[!ht]
    \centering
        \includegraphics[totalheight=12.5cm]{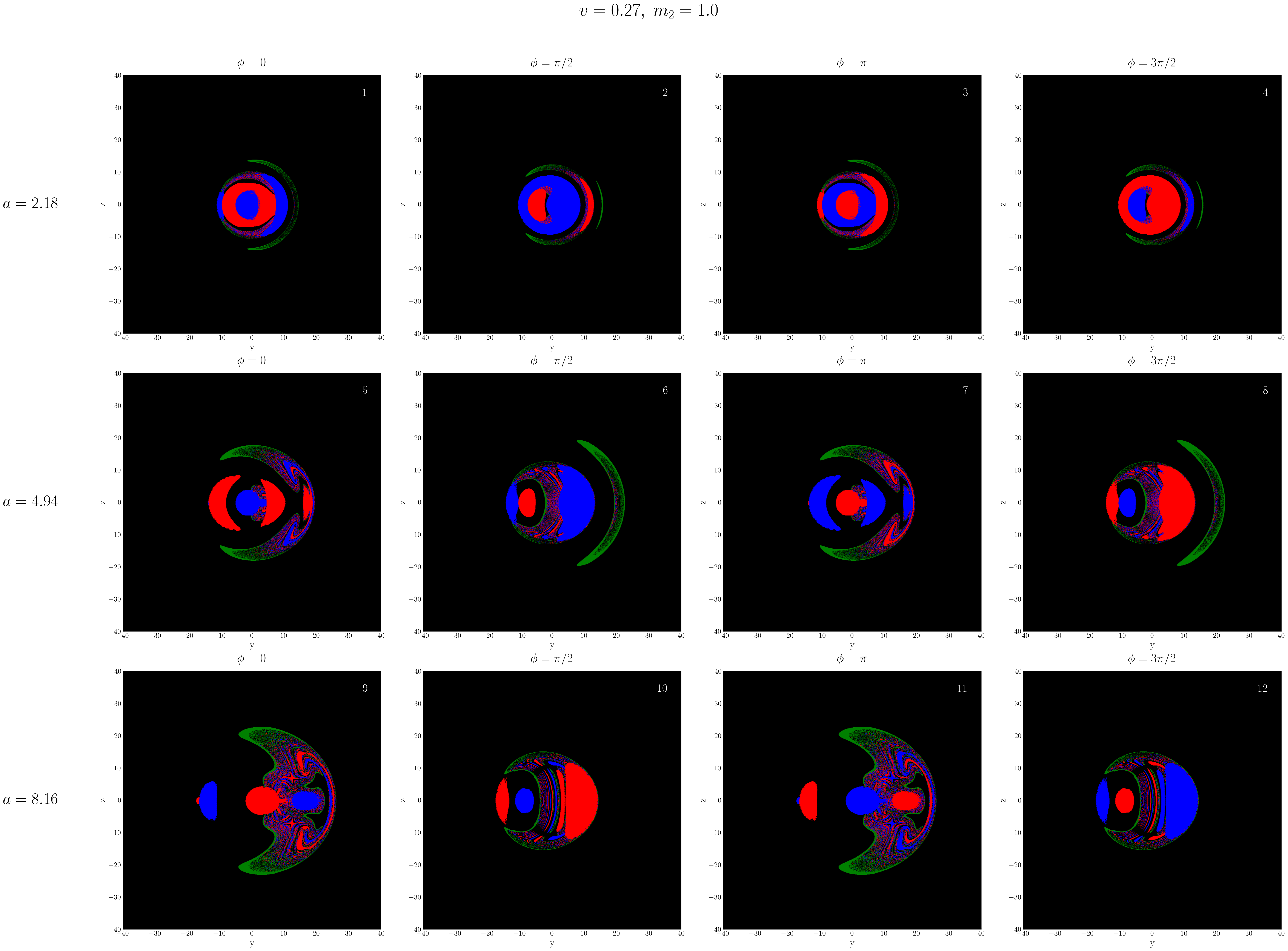}        
        \caption{Here we show the portrait for test particles of low incident velocity $v=0.27$, three semi-major axes, $a=2.18$, $a=4.94$, $a=8.16$ and four initial $\phi$. Points that exit the simulation before it finishes are colored black, those that impact the primary body are colored red, those that impact the secondary are colored blue and those that do not impact either body or exit, are colored green. All simulations are shown on the same scale} 
        \label{basins3}
    \end{figure}

    Finally we consider the portraits in the low-velocity case, shown in Fig(\ref{basins3}), which have an even more complicated structure. At small separations the effects of gravity strongly influence the dynamics, however, since the two bodies are so close together, the resulting potential is overall similar to that of a single body , except for a thin island of trajectories that become cooribital (subplots 1,2,3,4 in Fig(\ref{basins3}).  As the separation increases, this island becomes larger and begins to include a complex pattern of particles interacting with the primary and secondary.  Meanwhile, the central island distorts and shrinks and begins to feature points that become coorbital (subplots 5,6,7,8 in Fig(\ref{basins3}). The most complex situations are those with low velocities and large separations. In these cases, the island grows larger, dominating over the main shadow and threatening to enclose it. Uniquely among the simulations we perform, here complicated dynamical behavior within this island dominates the cross section (subplots 9,10,11,12 in Fig(\ref{basins3}).

\section{Mass Ratio Evolution of a Growing Binary}

    We can examine the role of our two body cross sections by considering the growth of a fiducial binary with an arbitrary mass ratio.  We consider a simplified case in which the two bodies grow solely due to an accretion process.  In this limit, the masses of  the bodies over time are described by

    \begin{equation}
        \frac{d M_i(t)}{d t} = \pi R_i^2 \left( 1 + \theta_i \right) v.
    \end{equation}

\begin{figure}[!ht]
    \centering
        \includegraphics[totalheight=5.5cm]{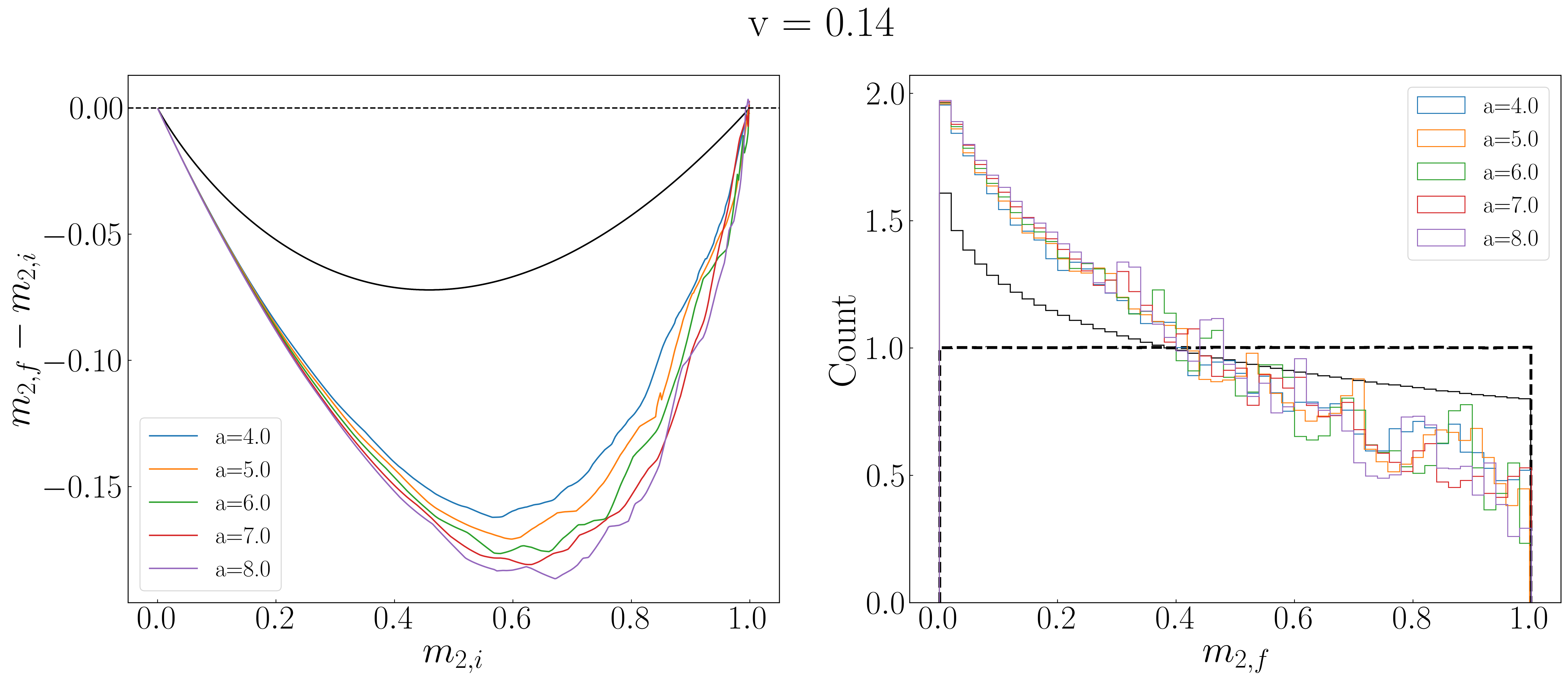}        
        \caption{In the left panel we show the change in mass ratio as a function of the initial mass ratio for an accreting binary using single-body cross sections for both objects (black) and our two-body cross sections for various initial semi-major axes (colored lines). In the right panel, we show the effect of these cross sections on the initially uniformly distributed population of binaries (dashed line).  The single-body cross sections (black) produce a less sharply peaked distribution than the two-body cross sections (colors)} 
        \label{fig:growth}
    \end{figure}

    Here the subscript $i$ indicates either the primary or secondary body.  We compare two scenarios. In the first, the focusing factor for both bodies is set to the naive estimate assuming the bodies are independent of each other, i.e. $\theta_i = \fracbrac{v_{esc,i}}{v}^2$.  In the second, the gravitational focusing factor is interpolated from our grid of simulations.  To simulate the proto-Kuiper belt, we consider a situation with significant gravitational focusing and choose $v=0.1$.  This is just a fiducial choice since our grid of cross sections enables similar calculations to be done over a variety of different $v$.  We examine the change in the mass ratio over the time it takes the primary to double in mass.  The results are shown in Fig(\ref{fig:growth}).  

    In both cases and for all initial mass ratios, the mass ratio shrinks as the bodies grow.  This occurs since, in all regimes at our chosen $v$, the larger mass has a proportionally larger cross section than the smaller mass.  In the single body case, the source of this effect is immediately clear as $\theta \sim v_{esc}^2$.  In the two-body case, the situation is a more complicated function of the secondary benefiting from the gravitational focusing of the primary, and the enhancement of the primary cross section due to the secondary's presence.  The right panel of Fig(\ref{fig:growth}) shows that not only does the enhancement of the primary cross section dominate in the two body cases, but it does so to a greater degree than would be expected from treating the two bodies separately. In addition, there is a semi-major axis dependent shift in where the largest change in mass ratio occurs, moving from $m_{2,i} \approx 0.5$ in the single body case to $m_{2,i} \approx 0.7$ in the case of bodies initially spaced by $8$ primary radii. 

    The center panel of Fig(\ref{fig:growth}) shows the effect of this change on an initially uniform population population of binaries of varying mass ratios. In the single-body case, there is a slight deficit of high-mass ratio bodies and an excess of those at smaller mass ratios. The two-body case results in a more extreme distribution with a larger excess of low mass ratio binaries, which occurs nearly independently of the chosen semi-major axis.  Since cross sections are enhanced in this regime relative to what would be expected in the single-body case, a modest amount of accretion will quickly shift the mass-ratio distribution towards smaller values.  
    
\section{Data Availability}

To facilitate future exploration of the topic we make all of our computed cross sections available at DOI: \href{https://doi.org/10.5281/zenodo.17329287}{10.5281/zenodo.17329287}. Or on Github at \href{https://github.com/r-zachary-murray/crosssections}{}

\section{Conclusion}

Currently, the distribution of binary mass ratios in the Kuiper belt is poorly characterized, due to both the small number of binaries detected and observational bias towards objects of low magnitude difference and thus large mass ratio. However, in the future LSST and the Rubin Observatory are poised to discover tens of thousands of additional Kuiper belt objects.  The mass ratio distribution of these binaries will be critical for assessing the formation mechanism of subsequent processes that produced the belt as we see it today.  We find that the relationship of the cross sections of an accreting binary is a complicated function of its orbital configuration.  The behavior of such a binary differs qualitatively from that of a single body, as the presence of the second body results in coorbital states and additional free variables. Numerically, we show these dynamics can result in an enhancement of the cross section by as much as a factor of $3$ above the single-body estimate or can shrink the cross section by as much as a factor of $2$ in other cases. These cross sections imply that growing binaries tend towards smaller mass ratios more quickly than would be expected given the single-body cross sections.  Future characterization of the mass distribution of the Kuiper belt, enabled by LSST, will allow a more detailed picture of the role of accretion in the natal Kuiper belt. 

\section{Acknowledgments}
We wish to thank Matt Holman for his input on the draft of this paper. 
This work was supported by the French government through the France 2030 investment plan managed by the National Research Agency (ANR), as part of the Initiative of Excellence of Université Côte d’Azur under reference number ANR-15-IDEX-01.

\bibliography{refs}{}

\end{document}